\documentclass[fleqn]{annalen}
\usepackage{pictex}
\usepackage{graphics}
\usepackage{epsf}
\begin{document}
\newcommand{\volume}{9}              
\newcommand{\xyear}{2000}            
\newcommand{\issue}{1}               
\newcommand{\recdate}{01.07.2002}    
\newcommand{\revdate}{dd.mm.yyyy}    
\newcommand{\revnum}{0}              
\newcommand{\accdate}{dd.mm.yyyy}    
\newcommand{\coeditor}{ue}           
\newcommand{\firstpage}{1}           
\newcommand{\lastpage}{5}            
\setcounter{page}{\firstpage}        
\newcommand{\ba}[1]{\begin{eqnarray} \label{(#1)}}
\newcommand{\ea}{\end{eqnarray}}
\newcommand{\be}[1]{\begin{equation} \label{(#1)}}
\newcommand{\ee}{\end{equation}}
\newcommand{\nn}{\nonumber}
\def \znbb {$0\nu\beta\beta~$}
\def \tnbb {$2\nu\beta\beta$}
\def \Rpv{R_{P} \hspace{-0.9em}/\;\:\hspace{0.8em}}
\def \emass {\langle m_{\nu} \rangle}
\def\lsim{\mathrel{\vcenter{\hbox{$<$}\nointerlineskip\hbox{$\sim$}}}}
\def\gsim{\mathrel{\vcenter{\hbox{$>$}\nointerlineskip\hbox{$\sim$}}}}
\def\ol{\overline}
\def\ul{\underline}
\def\half{\frac{1}{2}}
\def\quarter{\frac{1}{4}}
\def\third{\frac{1}{3}}
\def\thalf{\tfrac{1}{2}}
\def\tquarter{\tfrac{1}{4}}
\def\tthird{\tfrac{1}{3}}
\def\tsixth{\tfrac{1}{6}}
\def \Rp{$\not{\hspace*{-1mm}R}_p$}
\def\egzk{E_{\rm GZK}}
\def\tentwenty{10^{20}}
\def\nue{\nu_e}
\def\nunote{\nu_{\not e}}
\def\numu{\nu_{\mu}}
\def\nutau{\nu_{\tau}}
\def\nuebar{\bar{\nu}_e}
\def\numubar{\bar{\nu}_{\mu}}
\def\nutaubar{\bar{\nu}_{\tau}}
\def\dmsq{\delta m^2}
\def\dmatm{\delta m^2_{\rm atm}}
\def\dmsun{\delta m^2_{\rm sun}}
%
\newcommand{\keywords}{Neutrino mass, beyond the standard model, 
unification} 
\newcommand{\PACS}{12.60.-i, 14.60.Pq}
\newcommand{\shorttitle}{H.\ P\"as, Neutrino masses and physics beyond 
the standard model} 
\title{Neutrino masses and particle physics beyond the\\ standard model}
\author{Heinrich P\"as} 
\newcommand{\address}
  {Institut f\"ur Theoretische Physik und Astrophysik\\ Universit\"at 
W\"urzburg\\ Am Hubland, 97074 W\"urzburg, Germany}
\newcommand{\email}{\tt paes@physik.uni-wuerzburg.de} 
\maketitle
\begin{abstract}
The evidence for non-vanishing neutrino masses from
solar and atmospheric neutrinos provides the first 
solid hint towards physics beyond the standard model. 
A full reconstruction of the neutrino spectrum may well provide 
a key to the theoretical structures underlying the standard model
such as supersymmetry, grand unification or extra space dimensions.
In this article we discuss the impact of absolute neutrinos masses on
physics beyond the standard model. We review the 
information obtained from neutrino oscillation 
data and discuss the prospects of the crucial determination of the absolute
neutrino mass scale, as well as 
the intriguing connection with the Z-burst 
model for extreme-energy cosmic rays.
\end{abstract}

\section{Introduction}

The ultimate goal of particle physics is the quest for unification in a final 
theory underlying the standard model (SM), 
which describes the present knowledge
about physics at low energies.
So far the only experimental hint for new physics 
beyond the SM has been
provided by solar and atmospheric neutrino experiments which have 
established solid evidence for non-vanishing neutrino masses. 
This article aims to be a pedagogical introduction to the specific role 
of the neutrino among the elementary fermions and what can be learned 
from future neutrino experiments about the theoretical structures underlying 
the SM (for more detailed excellent reviews on the topic see also
\cite{conrev,reviews}). 
A crucial ingredient for the reconstruction of theories beyond the SM
will be a determination of 
the absolute mass scale for
neutrinos, which is still unknown \cite{pw}.
In fact, it is an experimental challenge
to determine an absolute neutrino mass below 1~eV.
Three approaches 
have the potential to accomplish the task, namely  
larger versions of the tritium end-point distortion measurements,
limits from the evaluation of the large scale structure of the universe,
and 
next-generation neutrinoless double beta decay ($0\nu\beta\beta$) experiments.
In addition, there is a fourth possibility: 
the extreme-energy cosmic-ray experiments 
in the context of the recently emphasized Z-burst model.
 
This article is organized as follows:
In section 2 the basic ideas of  supersymmetry, grand unification,
and large extra dimensions are 
mentioned. Section 3 deals with 
the specific role of the neutrino among the elementary fermions
of the SM,
as well as with the two most popular mechanisms for neutrino 
mass generation.  Also, 
the link of absolute neutrino masses to the theory 
underlying the SM is discussed. Section 4 reviews the experimental
evidence for non-vanishing neutrino masses from neutrino oscillations and
the knowledge about the neutrino mass matrix.
Section 5
deals with direct 
determinations of the absolute neutrino mass via tritium beta decay 
and cosmology.
In section 6 we discuss the $0\nu\beta\beta$  
which may test very small values of neutrino masses, when  
information is input 
obtained from oscillation studies. Section 7 finally 
deals with the connection of the sub-eV neutrino mass scale and the ZeV 
energy scale of extreme energy cosmic rays in the Z-burst model. 

\section{The holy grail of particle physics}
The final goal of particle physics is the unification of particles
and forces in a fundamental framework. The low energy particle content
of the SM consists of three families (flavors) of quarks
\be{}
\left(\begin{array}{c}u_L\\d_L\end{array}\right),~~
\left(\begin{array}{c}c_L\\s_L\end{array}\right),~~
\left(\begin{array}{c}t_L\\b_L\end{array}\right),~~
u_R,~d_R,~c_R,~s_R,~t_R,~b_R
\ee
and leptons
\be{}
\left(\begin{array}{c}e_L\\ \nu_e\end{array}\right),~~
\left(\begin{array}{c}\mu_L\\ \nu_\mu\end{array}\right),~~
\left(\begin{array}{c}\tau_L\\ \nu_\tau\end{array}\right),~~
e_R,~\mu_R,~\tau_R.
\ee
The left handed fermions transform as doublets under the 
electroweak SU(2) gauge group, the right handed counterparts are singlets,
and no right-handed neutrinos are included in the SM.
To preserve 
gauge invariance, fermion masses are understood as symmetry breaking 
effects due to couplings to the vacuum expectation value (vev)
of the Higgs 
doublet - in analogy to the effective photon masses generated
in superconductors.

\begin{figure}
\vspace{0.5cm}
\centerline{\resizebox{6cm}{6cm}{\includegraphics{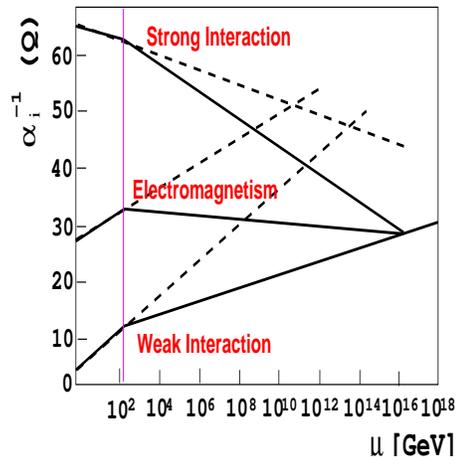}}}
 \caption{Running coupling constants: With increasing energy the couplings
of the electromagnetic, strong and weak interactions evolve, are supposed to
change directions
at the SUSY scale of about 1 TeV, and are finally assumed to
unify at the unification scale.}
 \label{fig:unif}
\end{figure}

The effective interactions at low energies are electromagnetism, weak
and strong interactions which are mediated by gauge bosons.
In the sense of unification, 
the running of the coupling strengths, 
due to vacuum polarization effects of
these interactions, are traced to higher energies and 
unification is assumed at the 
grand unified theory (GUT) scale (see Fig.~\ref{fig:unif}). 
In this picture, physics at
low energies is described by the 
SM, around 1 TeV supersymmetry enters, and at some $10^{16}$~GeV
unification is realized. Large extra space dimensions may complement this 
picture and can, in some respects, 
be considered as an alternative to low energy supersymmetry.

\subsection{Supersymmetry}
Supersymmetry (SUSY) is a hypothetical symmetry between bosons and 
fermions in the sense that
each elementary fermion $f$
acquires a superpartner, a scalar fermion (sfermion) $\tilde{f}$,
and each gauge or Higgs boson gets a spin 1/2 gaugino or 
Higgsino partner.
The inclusion of these new degrees of freedom at the SUSY breaking scale
of about 1 TeV allows the gauge couplings to unify. Supersymmetry 
would also cancel
divergencies of the SM and would offer a candidate for 
the spurious dark matter of the universe, provided 
the lightest supersymmetric 
particle is stable. Moreover, it would offer 
a natural and promising approach for developing
a consistent quantum theory of gravity. 

\subsection{Grand Unification}
At the unification scale, the three interactions are assumed to unify 
and the elementary fermions to be accommodated in large 
multiplets. In this way a higher level of symmetry is restored in the theory.
Typical issues are lepton and baryon (quark) 
number violation due to transitions between leptons and quarks 
in the same multiplet and the prediction of right handed neutrinos.

\subsection{Extra Space-Dimensions}
Extra dimensions beyond the three space and one time dimensions 
of the low energy world are predicted
in string theories which aim at a consistent quantum theory of gravity.
Such extra dimensions can be large (up 0.1~mm size, for example) if the
fermions of the SM are confined on a three dimensional brane
and only gravitons and SM singlets can propagate in the extra-dimensional bulk.
Such theories have interesting predictions for the energy scales of 
unification, the quantum gravity scale, and neutrino physics.

\section{Neutrino masses and physics beyond the standard model}
The specific role of the neutrino among the elementary fermions of the 
SM is twofold: It is the only neutral particle, and its mass
is much smaller than the masses of the charged fermions. Thus 
these properties may be related within a deeper theoretical 
framework underlying the SM -- usually via Majorana 
mass-generating mechanisms. Such lepton number violating
Majorana masses connect particle and 
antiparticle degrees of freedom and are thus obviously forbidden for 
charged particles.  
In the following, we comment on the most popular mechanisms to generate 
small neutrino masses, namely the see-saw mechanism, radiative neutrino 
mass generation and large extra dimensions.

\begin{figure}
\vspace{0.5cm}
\centerline{\resizebox{6cm}{4cm}{\includegraphics{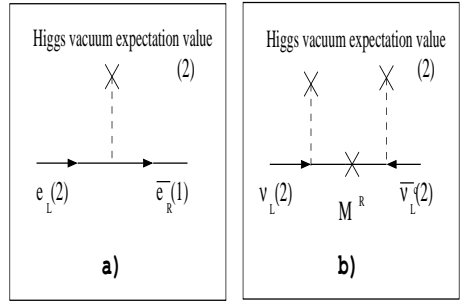}}}
 \caption{Diagrams for Dirac masses via couplings to the Higgs vacuum 
expectation value (panel a) and effective Majorana neutrino masses via
couplings to the Higgs vev and a heavy right-handed neutrino singlet
(panel b).
}
 \label{fig:massdias}
\end{figure}

\subsection{The see-saw mechanism}
\begin{figure}
\vspace{0.5cm}
\centerline{\resizebox{7cm}{7cm}{\includegraphics{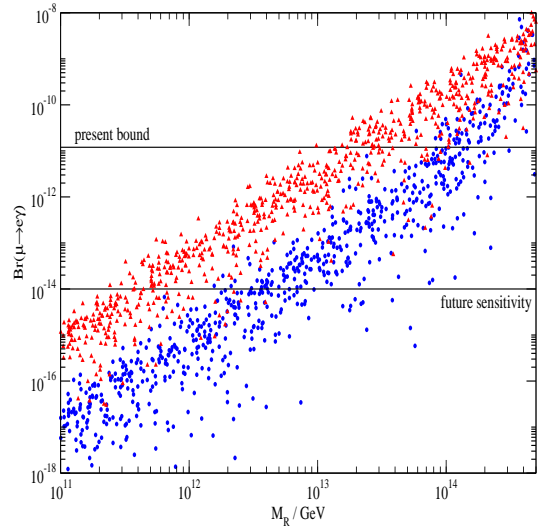}}}
 \caption{Computer simulation of the branching ratio 
$\mu \rightarrow e \gamma$ as a function of
the right-handed Majorana mass $M^R$. Shown are large (lower curve)
and small (upper curve) neutrino mass scales, for a specific SUSY (mSUGRA) 
scenario. The scatter points correspond to estimated uncertainties in neutrino 
parameters after planned neutrino experiments have been performed
(from \protect{\cite{dprr}}).}
 \label{depp}
\end{figure}

The see-saw mechanism \cite{seesaw}
is based on the observation that in order to generate
Dirac neutrino masses 
\be{}
m^D~ \overline{\nu^{}_L} \nu^{}_R
\label{dirac}
\ee
analogous to the mass terms of the charged leptons 
(see Fig.~\ref{fig:massdias}a), the 
introduction of right-handed SU(2) singlet neutrinos is required. 
However, a lepton-number violating,
Majorana mass term 
for right-handed neutrinos
\be{}
\overline{\nu_R}~ M^R~ (\nu_R)^c
\label{maj}
\ee
is not prohibited by any gauge
symmetry of the SM. Thus by buying 
a Dirac neutrino mass term $m^D$, one inevitably invites mixing with a  
Majorana mass $M^R \gg m^D$ 
which may live, a priori, at the mass scale of the underlying unified theory.
The diagonalization of the general mass matrix yields mass eigenvalues
\ba{}
m_{\nu} \simeq (m^D)^2/M^R \ll m^D, 
\label{seesaw}
\\
M \simeq M^R, 
\ea
explaining the smallness of the light mass, which can be described by
the effective operator corresponding to the diagram in 
Fig.~\ref{fig:massdias}b.
The fundamental scale $M^R$ is unaccessible for any kind of direct 
experimental testing. Nevertheless,
it is obvious from eq. (\ref{seesaw}) 
that with information on the low-energy observables 
$m_{\nu}$ and $m^D$ the ``beyond the SM" mass scale of $M^R$ can be 
reconstructed. While it turns out to be unrealistic to determine $m^D$ in 
the SM, this option exists indeed in supersymmetry.
In the  supersymmetric version of the see-saw mechanism,
lepton flavor violation (LFV) in the neutrino mass matrix 
(as required by neutrino oscillations) generates 
also LFV soft terms in the slepton mass matrix
proportional to the Dirac neutrino
Yukawa couplings \cite{hisano},
\be{}
\delta \tilde{m}_L^2 \propto Y_{\nu} Y_{\nu}^{\dagger}.
\ee
These LFV soft terms induce large branching ratios for SUSY mediated 
loop-decays such as 
$\mu \rightarrow e \gamma$,
\be{}
\Gamma(\mu \rightarrow e \gamma) \propto
\alpha^3 \frac{|(\delta \tilde{m}_L)^2_{ij}|^2}{m_S^8} \tan^2 \beta\,.
\ee
Here $m_S$ denotes the slepton mass scale in the loop.
Thus in the supersymmetric framework it is possible to probe the heavy mass 
scale $M^R$ by determining the (light) neutrino mass scale $m_\nu$
and the (Dirac) Yukawa couplings. This fact is illustrated in 
Fig.~\ref{depp}, where computer simulation data of the branching ratio 
$\mu \rightarrow e \gamma$ 
are shown as a function of $M^R$
for a specific SUSY (mSUGRA) scenario, both for a large (lower curve)
and small (upper curve) neutrino mass scale \cite{dprr}. 

\subsection{Radiative neutrino masses}

\begin{figure}
\vspace{0.5cm}
\centerline{\resizebox{8cm}{4cm}{\includegraphics{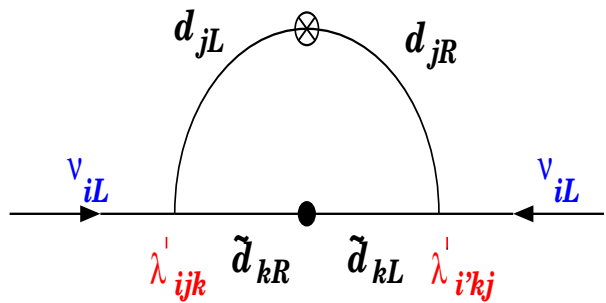}}}
 \caption{Radiative generation of neutrino Majorana masses in \Rp-violating
SUSY.}
 \label{loop}
\end{figure}

An alternative mechanism generates neutrino masses via loop graphs at the
SUSY scale, in contrast to the tree level generation of charged lepton masses
via the Higgs mechanism (see e.g. \cite{val}).
In supersymmetric theories lepton-number violating couplings,
$\lambda$ and $\lambda'$, may arise if the discrete 
R-parity symmetry is broken (\Rp). These couplings may
induce neutrino masses via one loop self-energy graphs, see Fig.~\ref{loop}.
The entries in the neutrino mass matrix, given by
\be{}
m_{\nu_{ii'}} \simeq {{N_c \lambda'_{ijk} \lambda'_{i'kj}}
\over{16\pi^2}} m_{d_j} m_{d_k}
\left[\frac{f(m^2_{d_j}/m^2_{\tilde{d}_k})} {m_{\tilde{d}_k}} +
\frac{f(m^2_{d_k}/m^2_{\tilde{d}_j})} {m_{\tilde{d}_j}}\right],
\ee
are proportional to the products of \Rp-couplings and depend on the 
values of superpartner masses.
A determination of the absolute neutrinos mass scale would allow one to 
constrain all entries in the mass matrix, using the smallness of 
atmospheric and
solar $\Delta m^2$'s and unitarity of the neutrino mixing matrix $U$.
In fact, recent bounds on absolute neutrino masses improve
previous bounds on \Rp-couplings by up to 4 orders of magnitude \cite{bkp}.
Thus determining the neutrino mass probes physics at heavy 
mass scales beyond the SM also in the case of radiatively 
generated neutrino masses.

\subsection{Neutrino masses from large extra dimensions}
In theories with large extra dimensions small neutrino masses may be generated
by volume-suppressed  couplings to right-handed neutrinos which can
propagate in the bulk, by the breaking of lepton number on 
a distant brane, or by the curvature of the extra dimension \cite{xtranus}.
Thus neutrino masses can provide information about the volume or 
the geometry of the large extra dimensions. 

\section{Experimental evidence: neutrino oscillations}

\begin{figure}
\vspace{0.5cm}
\centerline{\resizebox{10cm}{5cm}{\includegraphics{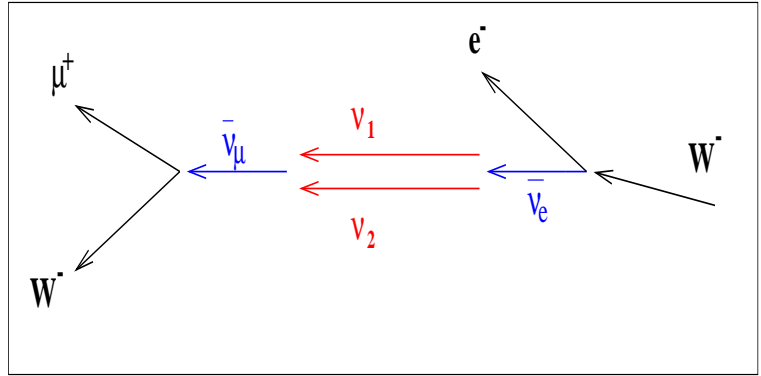}}}
 \caption{Schematic diagram for neutrino oscillations.}
 \label{osc}
\end{figure}

Neutrino oscillations, i.e. oscillating flavor conversion, 
arise if the flavor states $\nu_{\alpha}$
are superpositions of different non-degenerate mass eigenstates $\nu_i$,
\be{}
| \nu_\alpha \rangle = \sum_{i} U_{\alpha i} | \nu_i \rangle.
\ee
In this case, a flavor eigenstate produced at one vertex propagates as
a superposition of mass eigenstates (see Fig.~\ref{osc}),
\be{}
| \nu_{\alpha} \rangle = \sum_{i} e^{-i E_i t} U_{\alpha i} | \nu_i \rangle,
\ee
with energies $E_i=\sqrt{p_i^2 + m_i^2}$. At a second vertex, there is
the probability
\ba{}
P(\nu_\alpha \rightarrow \nu_{\beta})(t)&=& 
|\langle \nu_\beta| \nu_\alpha \rangle|^2 \nn \\
&=& | \sum e^{-i E_i t} U_{\alpha i} U^*_{\beta i}|^2
\ea
to observe a different flavor eigenstate $\nu_{\beta}\neq \nu_{\alpha}$.
In a two neutrino framework, $U$ can be parametrized as
\be{}
U=\left(\begin{array}{cc} \cos \theta & \sin \theta\\
-\sin \theta & \cos \theta \end{array} \right).
\ee 
The oscillation probability becomes
\be{}
P(\nu_\alpha \rightarrow \nu_{\beta})(t)= \sin ^2 2 \theta \sin^2 
\left[\frac{\Delta m^2}{4 E} x \right],
\label{eq:oscprop}
\ee
with the propagation distance $x \simeq t$ and 
$E_i \simeq |\vec{p}| + \frac{m_i^2}{2|\vec{p}|} \simeq |\vec{p}|$.
The two-neutrino approximation is a good approach to describe solar and 
atmospheric neutrino oscillations since the remaining mixing 
angle(s) are small.

\subsection{Reactor neutrinos}
An intense 
terrestrial source of low energy 
MeV $\bar{\nu}_e$ neutrinos is provided by nuclear
reactors. Thus reactor neutrino experiments searching for 
$\bar{\nu}_e \rightarrow \bar{\nu}_{\not e}$ oscillations offer a 
possibility to test small $\Delta m^2$'s (compare eq.~(\ref{eq:oscprop})). 
In the CHOOZ \cite{Apollonio:1999ae}
and Palo Verde \cite{Boehm:2001ik} experiments one didn't
 observe a disappearance
signal. One has restricted the mixing matrix element to the small value
of $U^2_{e3}<0.025$ or $\sin^2 2 \theta_{13}<0.1$, for 
$\Delta m_{13}^2>7 \times 10^{-4}$~eV$^2$. This bound is also important for
the interpretation of the atmospheric neutrino anomaly.

\subsection{Atmospheric neutrinos}
\begin{figure}
\vspace{0.5cm}
\centerline{\resizebox{5cm}{5cm}{\includegraphics{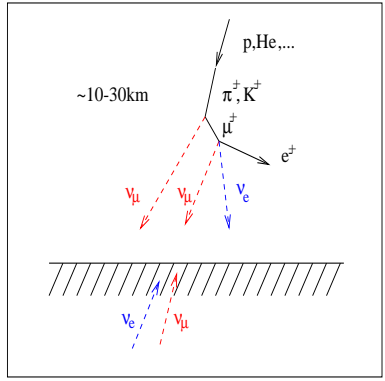}}}
 \caption{Atmospheric neutrino production from $\pi$ and $K$ decays.
Upward going neutrinos have propagated through the earth over a
distance of about $10^4$ km. This leads to enhanced oscillation
probablilities as compared to
neutrinos created above the detector.}
 \label{fig:atm}
\end{figure}

Atmospheric neutrinos are produced in the decays of the $\pi$ and $K$ mesons 
stemming from  cosmic ray primary reactions in the upper atmosphere.
Just by counting naively the neutrinos from the decay chain in 
Fig.~\ref{fig:atm},
a ratio of $(\nu_{\mu}+\overline{\nu}_{\mu})/(\nu_{e}+\overline{\nu}_{e})
\sim 2$ can be obtained for $E_{\nu}< 1$~GeV and 
$(\nu_{\mu}+\overline{\nu}_{\mu})/(\nu_{e}+\overline{\nu}_{e})
\gsim 2$ for $E_{\nu}> 1$~GeV (where not all muons decay before they reach
the detector). 

\begin{figure}
\centerline{\resizebox{7cm}{7cm}{\includegraphics{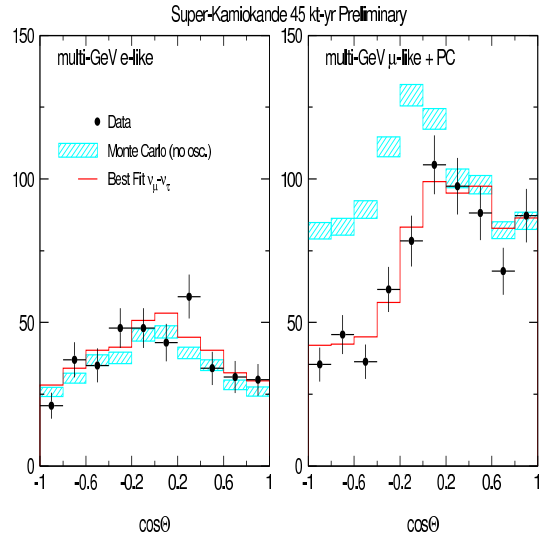}}}
 \caption{Zenith angle dependence of e-like and $\mu$-like events
in Super-K. The data show a clear zenith angle dependence for $\mu$-like
events while the e-like events behave as the no-oscillation expectation
\protect{\cite{superkatm}}.}
 \label{fig:atm2}
\end{figure}

Here the uncertainty in the flux is estimated 
to be $\sim$ 30 \%, while the uncertainty in the ratio is reduced to 
$\sim$ 5 \%. The following experimental observations provide a strong
indication for neutrino oscillations:

\begin{itemize}

\item
Reduced muon over electron neutrino
ratios by almost a factor of two
have been observed in the following experiments:
by the Soudan2 iron calorimeter, the 
MACRO liquid scintillator and the IMB, Kamiokande and 
the high statistics Superkamiokande (Super-K) \cite{superkatm}  
Cherenkov detector experiments. 
(These observations solved the issue of systematic errors in Cherenkov counters
caused by the fact that the early iron calorimeter experiments 
NUSEX and FREJUS did not observe a 
reduction.)
The reduced ratio implies either 
$\nu_{\mu}$ ($\bar{\nu}_{\mu})$ disappearance or 
$\nu_{e}$ ($\bar{\nu}_{e})$ appearance.

\item
The observed event rates at Super-K exhibit a zenith angle dependence. 
This reflects
the fact that upcoming neutrinos have propagated about $\sim 10^4$ 
km through the
earth, while neutrinos originating from the atmosphere above the detector
have propagated some 10-30 km, only (see Fig.~\ref{fig:atm}). The different 
propagation distances yield different oscillation probabilities.
The measured zenith angle spectra are shown in Fig.~\ref{fig:atm2}.

\item
The first data of the long baseline accelerator experiment K2K \cite{k2k} 
suggests confirmation of the 
atmospheric neutrino results.

\end{itemize}

Moreover oscillations of muon to electron neutrinos $\nu_e$
or additional ``sterile'' SU(2) singlet neutrinos $\nu_s$ can be excluded 
(at least to be the dominant process) from the data:

\begin{itemize}

\item
The CHOOZ and Palo Verde reactor experiments exclude 
$\nu_{\mu} \rightarrow \nu_e$ oscillations for the parameter range
of the atmospheric favored region with the non-observation of $\nu_e$ 
disappearance.


\item
In principle, sterile neutrinos imprint atmospheric data in three 
different ways: 
via differing matter effects for $\numu$ oscillation to 
$\nu_\tau$ vs.\ $\nu_s$,
via neutral current scattering of $\nu_{\tau}$ but not $\nu_s$, 
as measured by $\pi^0$ production;
and via $\tau$ appearance from $E_{\nu_{\tau}}$ scattering 
above threshold. The data show no evidence for these effects and this 
translates into bounds on the sterile neutrino component in atmospheric 
neutrinos.

\end{itemize}

In summary fits to the experimental results at different energies
and zenith angles single out $\nu_{\mu}\rightarrow \nu_{\tau}$
oscillations \cite{lpw} and a single region in the 
$\Delta m^2_{\rm atm}-\sin^2 2 \theta_{\rm atm}$ parameter space 
\cite{conrev,atmanal}, i.e. $1 \times 10^{-3}~{\rm eV}^2 <\Delta m^2_{\rm atm}
<4 \times 10^{-3}~{\rm eV}^2$, with best fit
$\Delta m^2_{\rm atm}=2.6 \times 10^{-3}$~eV$^2$,
and (close to) maximal mixing $\sin^2 2 \theta_{\rm atm}>0.7$, with best fit
$\sin^2 2 \theta_{\rm atm}=0.97$.

\subsection{Solar neutrinos}
\begin{figure}
\centerline{\resizebox{8cm}{8cm}{\includegraphics{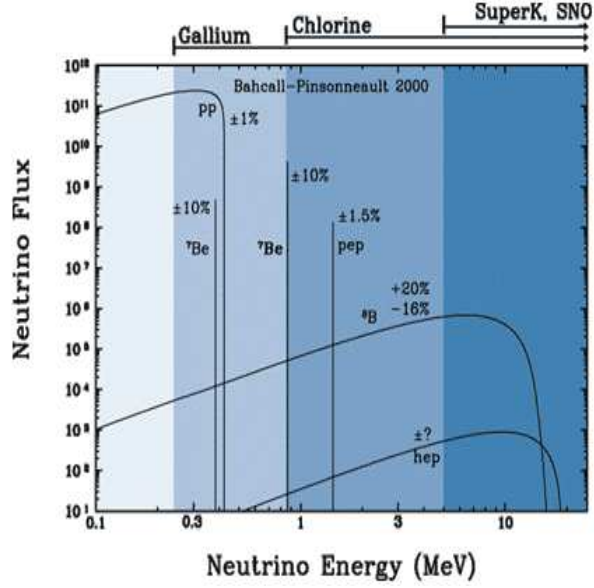}}}
 \caption{Solar neutrino spectrum: Fluxes as a function of energy for 
the different production processes (from \protect{\cite{bahcallweb}}). 
}
 \label{solspek}
\end{figure}

\begin{figure}
\centerline{\resizebox{8cm}{8cm}{\includegraphics{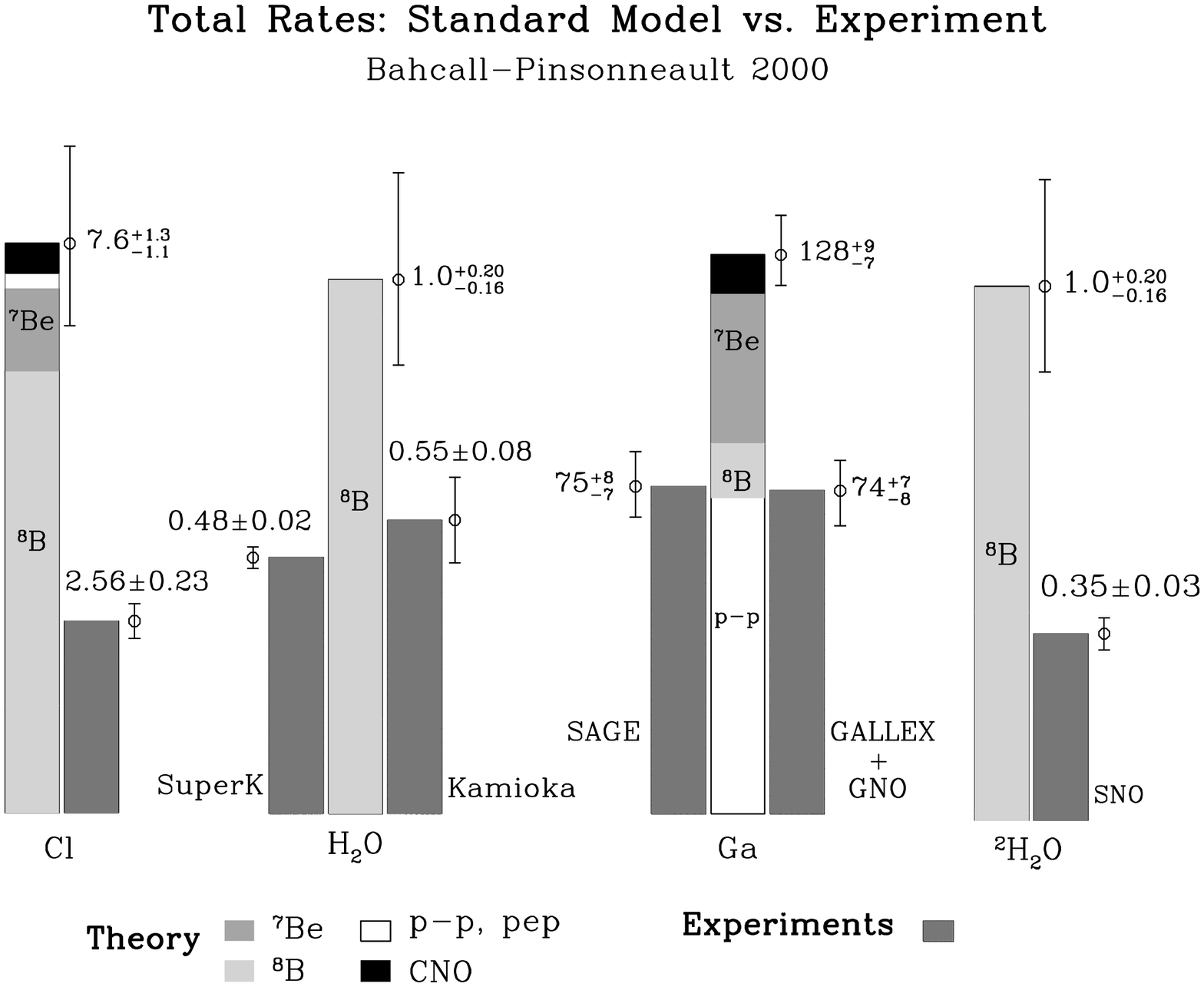}}}
 \caption{Measured rates of neutrino experiments versus expectations 
in the solar standard model (from \protect{\cite{bahcallweb}}). 
}
 \label{thvsexp}
\end{figure}

Solar neutrinos are produced by the nuclear fusion reactions in the 
solar core. The most important neutrino sources are the reactions of the pp
cycle
\ba{}
p+p &\rightarrow& D + e^+ +\nu_e, \\
p+e^- + p  &\rightarrow& D + \nu_e, \\
^{7}Be + e^- &\rightarrow& ^{7}Li + \nu_e, \\
^{8}B &\rightarrow& ^{8}Be^* + e^+ + \nu_e,
\ea
which give rise to the $pp$, $pep$, $^{7}Be$ and $^8B$ neutrinos,
respectively. A small contribution to the fluxes is produced in the CNO cycle.
The corresponding energy spectra are displayed in 
Fig.~\ref{solspek}, 
together with the sensitivities of the different experiments
due to the different energy thresholds of the radiochemical Gallium 
(Gallex/GNO \cite{gallex} and Sage \cite{sage}), Chlorine \cite{homestake}, 
and Cherenkov (Super-K \cite{superksol}, 
Kamiokande \cite{kam} and SNO \cite{sno})
detectors. Fig.~\ref{thvsexp} compares the expected 
(no-oscillation) neutrino fluxes with the actual measurement for the 
different experiments. The following conclusions can be drawn: 

\begin{itemize}

\item
All experiments observe less neutrinos than expected according to 
the standard solar model (SSM) \cite{bp00}.

\item
The combination of different experiments exhibits an energy dependent
suppression of solar neutrino fluxes. If $^8B$ neutrinos are observed at 
Super-K and SNO, there should be $pp$ and $^7Be$ neutrinos in the Gallium 
experiments as well. The measured rate, however, allows just for the
(solar-model independent) $pp$ neutrino flux, without any room left for 
contributions from $^{7}Be$ neutrinos (the absence of $^{7}Be$ neutrinos will
be tested by the BOREXINO \cite{borex} experiment, which is scheduled to start 
taking data in 2002).

\item
The SNO experiment is sensitive on both neutral current 
(due to any neutrino flavor) and charged current (due to electron 
neutrinos only)
reactions via the dissociation of deuterium. 
While the neutral current data are in
perfect agreement with the fluxes predicted of the SSM, 
$\Phi_{NC}/\Phi_{SSM}=1.01 \pm 0.12$,
the charged current
data show a suppression by roughly a factor of three.
This observation can be interpreted as evidence for the appearance of 
solar neutrinos with a non-electron flavor.

\end{itemize}

Matter effects play 
a crucial role in the interpretation of the solar neutrino results. 
The Mikheyev-Smirnov-Wolfenstein (MSW)  
effect \cite{msw} is a result of different matter potentials. Larger
effective masses are generated in the sun for electron 
neutrinos (which interact via neutral and charged current)
as compared to other flavors (which interact via neutral current only). 

If the (i) vacuum mass (squared) difference $\Delta m_{\odot}^2$
is not too large, a level crossing arises
in the sense that the heavier mass eigenstate in the sun is the 
electron neutrino $\nu_e$, while in vacuum it is mainly a different 
flavor $\nu_x$. 

For an (ii) adiabatic transition out of the solar core, 
the electron neutrino created
in the sun is converted resonantly into the heavy mass eigenstate in vacuum,
being mainly $\nu_x$.

This process 
is especially effective for (iii) not too large mixing, since for small 
mixing an almost pure $\nu_e$ state is converted into an almost pure $\nu_x$ 
state. 

The MSW conditions (i), (ii), and (iii) define isocontour  
lines for
$P(\nu_e \rightarrow \nu_{\not{e}})$ of triangle shape
in the $\Delta m_{\odot}^2-\sin^2 2 \theta_{\odot}$ plane. 
The boundaries of these isocontour lines are at (i) large $\Delta m^2$'s,
(ii) small $\Delta m^2$ combined
with small mixing angles and (iii) large 
mixing angles. Since any experiment gives a range of probabilities because of
finite error bars, the allowed regions are bands, whose limiting curves have 
the shape of a triangle. 
The superposition of the bands corresponding to different experiments 
defines through the overlap 
favored regions usually known as the small (SMA) and large (LMA) mixing 
angle MSW solutions. 

The SMA solution has been disfavored by the flat energy spectrum measured at 
Super-K. After the release of the SNO data, the LMA solution is 
selected at a 99 \% confidence level (C.L.) 
and is also favored if one only analyzes the total rates.
Also solutions with $\theta_{\odot}\geq \pi/4$ have been excluded.
Thus the solution to the solar neutrino problem can be summarized as
\cite{solanal} $\nu_{e} \rightarrow \nu_{\mu,\tau}$ oscillations
with $2.7 \times 10^{-5}~{\rm eV}^2 <
\Delta m^2_{\odot}< 1.8 \times 10^{-4}~{\rm eV}^2$, with best fit
$\Delta m^2_{\odot}=5.6 \times 10^{-5}$~eV$^2$, and 
$0.27 < \tan^2 \theta_{\odot}< 0.55$, with best fit
$\tan^2 \theta_{\odot}=0.39$. 
It has become customary to express a mixing angle
sensitive to matter effects, such as the solar angle,
as $\tan\theta$ rather than $\sin2\theta$
in order to account for the octant of the ``dark side'',
$\pi/4 <\theta\leq\pi/2$.

Limits on the solar mode $\nue \rightarrow \nu_s$ result from 
model fits to the Super-K solar data, but especially
from the recent SNO data.
There is no evidence favoring a sterile admixture in the 
neutrino flux from the sun.  
Nevertheless, a large sterile component 
remains compatible with the SNO result, 
because of the uncertainty in the true high-energy solar 
neutrino flux produced by the $^8$B reaction in the sun.

\subsection{LSND and a 4th sterile neutrino?}

A third experimental evidence has been reported by the LSND experiment
\cite{lsnd}. LSND was searching for $\bar{\nu}_{\mu} \rightarrow \bar{\nu}_e$ 
and $\nu_{\mu} \rightarrow \nu_e$ appearance from the products
of pion and subsequent muon decay, produced by the scattering of accelerated 
protons on a fixed target:
\ba{}
p+{\rm target} &\rightarrow& \pi^{+} +X, \nn \\
&& \pi^+ \rightarrow \mu^+ \nu_{\mu}~~~~{\rm (decay~in~flight)}, \nn \\
&& \mu^+ \rightarrow e^+ \nu_e \bar{\nu}_{\mu} ~~~~{\rm 
(decay~at~rest)}.
\ea
LSND has observed a clear excess of events with $\bar{\nu}_e$
signature, which has been interpreted as evidence for neutrino 
oscillations with 
$P(\bar{\nu}_{\mu} \rightarrow \bar{\nu}_e)=0.31 \% (^{+0.11}_{-0.10}  
\% \pm 0.05 \%)$. 
In addition an analysis of $\nu_{\mu} \rightarrow \nu_e$ oscillations from the
decay in flight has been performed and an oscillation probability of
$P(\nu_\mu \rightarrow \nu_e)=0.26 \% (\pm 0.07 \% \pm 0.05 \%)$
has been deduced, being consistent with the 
$\bar{\nu}_{\mu} \rightarrow \bar{\nu}_e$ results. 
The favored regions
are constrained by the negative result of the KARMEN experiment, which is similar to LSND but possesses a smaller baseline.
A combined analysis favors
$\nu_{\mu} \rightarrow \nu_e$ oscillations with 
$0.12~{\rm eV}^2 <\Delta m^2_{\rm LSND} < 1.1$~eV$^2$ and 
$10^{-3}< \sin^2 2 \theta_{\rm LSND} < 0.7$.

Taken at face value, the solar, atmospheric, and LSND data 
require three independent $\Delta m^2$ scales.  
Thus, four neutrinos seem to be required.
The $Z$-boson width further requires that one of these
four neutrinos be a ``sterile'' $SU(2)\times U(1)$ electroweak-singlet.
In the so-called 2+2 spectrum, the LSND scale splits
two pairs of neutrino mass-eigenstates.
Phenomenologically, it is required that one pair mix $\numu$ with
$\nutau$ and $\nu_s$ for explaining the atmospheric $\numu$ disappearance,
while the second pair mix $\nue$ with $\nutau$ and $\nu_s$
for explaining the solar $\nue$ disappearance.  
The small LSND amplitude is 
accommodated with a small mixing of $\nue$ into the first pair,
and a small mixing of $\numu$ into the second pair.

\begin{figure}[!t]
\vspace{0.5cm}
\centerline{\resizebox{8cm}{8cm}{\includegraphics{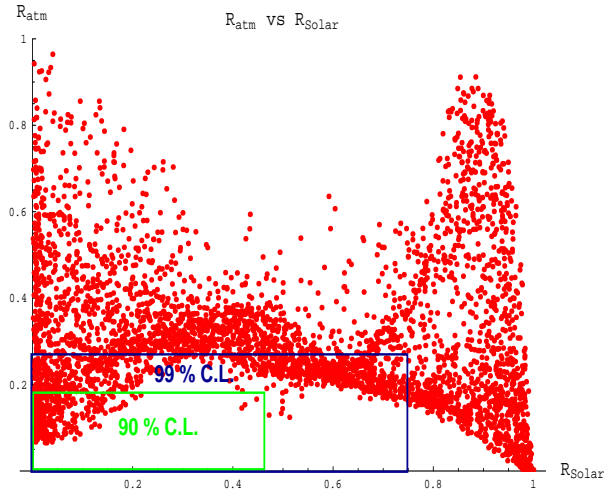}}}
 \caption{Sum rule for sterile neutrinos. 
The scatter points correspond to allowed values of small mixing angles.
Also shown are the allowed regions for $\nu_s$ admixture in solar and 
atmospheric neutrinos at 90 \% C.L. and 99 \% C.L.
(from \protect{\cite{song}}).}
 \label{fig:song}
\end{figure}

A sum rule requires a robust contribution of 
$\numu\rightarrow\nu_s$ in the atmospheric data and/or 
$\nue\rightarrow\nu_s$ in the solar data. 
The essence of the sum rule is that the sterile neutrino
may hide from solar oscillations, or from atmospheric oscillations,
but cannot hide from both \cite{Peres:2001ic}.
When the small angles, which mix the neutrinos in the atmospheric pair
with the ones in the 
solar pair, are neglected, the sum rule states that 
the probabilities to produce $\nu_s$ in the solar and in the atmospheric 
data sum to unity:
\be{SR1}
\left[\frac{P(\nu_e \rightarrow \nu_s)}{P(\nu_e \rightarrow \nu_{\not{e}})}
\right]_{\rm sun}
+\;
\left[\frac{P(\nu_{\mu} \rightarrow \nu_s)}{P(\nu_\mu \rightarrow \nu_{\not{\mu}})}
\right]_{\rm atm}
= 1.
\label{SR1}
\ee
  The relaxation of this sum rule, when 
matter effects and small angles are included, has been studied in \cite{song}
taking into account the bounds on $\nu_s$ admixture in 
solar and atmospheric neutrinos \cite{con2+2}.
In Fig.~\ref{fig:song} the approximate sum
rule is analyzed in generality, 
varying the usually neglected mixing angles in
their experimentally allowed ranges 
including possible matter effects.
The strong bound from atmospheric neutrinos is barely consistent with 
the 2+2 scheme, when 90 \% C.L. bounds are applied.
At 99 \% C.L., though, the 2+2 four neutrino 
scheme with sterile neutrinos is still allowed
(compare also the fit results in \cite{con2+2}).
The neutrino oscillation solution to LSND experiment will 
definitely be tested soon by the accelerator experiment MiniBooNE, which 
has an improved background and a significant larger baseline.

An interesting and elegant 
alternative to an oscillation solution for LSND is provided by lepton number 
violating anomalous muon decays, 
\be{}
\mu^+ \rightarrow e^+ \bar{\nu}_e \bar{\nu}_i,
\ee
which does not require sterile neutrinos
and not necessarily predicts a 
MiniBooNE signal \cite{bergmann}.

\begin{figure}
\vspace{0.5cm}
\centerline{\resizebox{12cm}{12cm}{\includegraphics{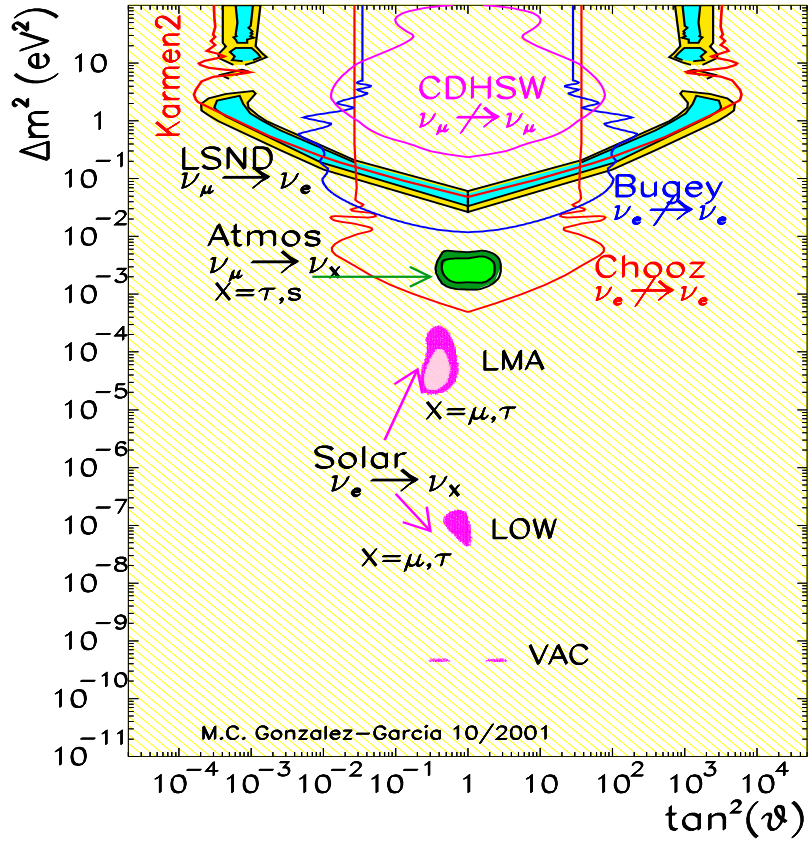}}}
 \caption{Summary of evidences for neutrino oscillations (from \cite{conrev}).}
 \label{summary}
\end{figure}

\subsection{Neutrino Oscillation Summary}
Over the past years a unique picture of neutrino mixing has been evolved,
see Fig.~\ref{summary}.
The bounds on sterile neutrinos from solar and atmospheric neutrino data
suggest strongly a three neutrino framework. Large/maximal mixing has been 
established for solar and atmospheric neutrinos, respectively, and the
third mixing angle is strongly bounded by reactor neutrino experiments.
The neutrino mixing matrix, modulo CP violating phases, turns out to be of
the following approximate form \cite{Barger:1998ta}:
\ba{}
U&=&\left(\begin{array}{ccc} 
\cos \theta_{\odot} & \sin \theta_{\odot} & 0\\
-\frac{\sin \theta_{\odot}}{\sqrt{2}} 
& \frac{\cos \theta_{\odot}}{\sqrt{2}} & \frac{1}{\sqrt{2}}\\  
\frac{\sin \theta_{\odot}}{\sqrt{2}} 
& \frac{-\cos \theta_{\odot}}{\sqrt{2}} & \frac{1}{\sqrt{2}}\\ 
\end{array} \right) 
\approx
\left(\begin{array}{ccc} 
\frac{1}{\sqrt{2}} & \frac{1}{\sqrt{2}} & 0\\
-\frac{1}{2} & \frac{1}{2} &  \frac{1}{\sqrt{2}}\\
\frac{1}{2} & -\frac{1}{2} &  \frac{1}{\sqrt{2}}
\end{array} \right) \nn \\
\ea
In addition, the mass squared differences 
$\Delta m^2_{\odot}\simeq 6 \times 10^{-5}$~eV$^2$ and
$\Delta m^2_{\rm atm} \simeq 3 \times 10^{-3}$~eV$^2$ are known. Future 
neutrino oscillation experiments will increase the accuracy of 
these parameters as follows:

\begin{itemize}

\item {\it $\Delta m_{\odot}^2$ and $\sin^2 2 \theta_{\odot}$:}
The long-baseline reactor experiment KamLAND is designed to test the LMA
MSW solution of the solar neutrino problem. Data taking has been started
in 2002 and the solar neutrino parameters will be determined with an accuracy
of 
$\delta(\Delta m_{\odot}^2)/\Delta m_{\odot}^2 = 10 \%$ 
and $\delta(\sin^2 2 \theta_{\odot})= \pm 0.1$
within three years of measurement
\cite{Barger:2001hy}.

\item {\it $\Delta m_{\rm atm}^2$ and $\sin^2 2 \theta_{\rm atm}$:}
The atmospheric oscillation parameters will be determined by the long-baseline 
accelerator experiment MINOS with an accuracy of\\ 
$\delta(\Delta m_{\rm atm}^2)/\Delta m_{\rm atm}^2 = 
30 \%$ and $\delta(\sin^2 2 \theta_{\rm atm})= \pm 0.1$
\cite{minos}.

\item {\it $\sin^2 2 \theta_{13}$:} 
The long baseline 
experiment MINOS \cite{minos} has a sensitivity down to $\sin^2 2 \theta_{13}<0.02$-$0.05$.
A future neutrino factory 
\cite{Barger:2000dy}
or the
analysis of the neutrino energy spectra of a future galactic supernova 
\cite{Dighe:2000bi} may 
provide a sensitivity down to $10^{-4}$.

\item {\it Direct/inverse type of hierarchy:}
The inverse hierarchical spectrum with two heavy and a single light state
is disfavored according to a recent analysis \cite{inv} of the neutrino 
spectrum from supernova SN1987A, unless the mixing angle $\theta_{13}$ is 
large (compare, however, \cite{Barger:2002px}). 

\end{itemize}

Nothing is known so far about the remaining parameters, the three CP 
violating phases (one Dirac and two Majorana phases) and the absolute 
neutrino mass scale.

\begin{itemize}

\item
At a neutrino factory one may be able to 
distinguish \(\delta = 0\) from $\pi/2$ if 
$\Delta m^2_{\odot}>10^{-5}$~eV$^2$
\cite{Gomez-Cadenas:2001ev}. 

\item
Even more difficult is the determination of 
Majorana phases, some information might be obtained by comparing
positive results in future double beta decay and tritium beta decay 
experiments \cite{phases}
or by testing related sneutrino mass matrices in supersymmetric 
theories.

\end{itemize}

Especially important for obtaining information on the theoretical structures
underlying the SM of particle physics is the full reconstruction 
of the neutrino mass spectrum and thus the
determination of the absolute neutrino mass.
The rest of this review concentrates on this difficult task.

\section{Absolute neutrino masses: direct determination}

While neutrino oscillation experiments provide information on the
neutrino mass squared differences $\Delta m^2_{ij}$, the absolute scale of 
the neutrino masses is not known so far. Upper bounds can be obtained from 
the neutrino 
hot dark matter contribution to the cosmological large scale structure
evolution and the Cosmic Microwave Background, from the interpretation 
of the extreme energy cosmic rays in the Z-burst model, from
tritium beta decay experiments, and, most importantly, from neutrinoless 
double beta decay experiments 
\cite{pw}.

\subsection{Tritium beta decay}

In tritium decay, the larger the mass states comprising $\nuebar$,
the smaller is the Q-value of the decay.
The manifestation of neutrino mass is a reduction of phase space
for the produced electron at the high energy end of its spectrum.
An expansion of the decay rate formula about $m_{\nue}$ leads to
the end point sensitive factor 
\be{}
m^2_{\nu_e}\equiv \sum_j\,|U_{ej}|^2\,m^2_j\,,
\ee
where the sum is over mass states which can kinematically alter
the end-point spectrum.
If the neutrino masses are nearly degenerate,
then unitarity of $U$ leads immediately to a bound on the heaviest 
neutrino mass eigenstate
$\sqrt{m^2_{\nu_e}}=m_3$.
The design of a larger tritium decay experiment (KATRIN) for improving 
the present 2.2~eV $m_{\nu_e}$ bound is under discussion;
direct mass limits as low as 0.4~eV, or even 0.2~eV, may be possible
in this type of experiment \cite{katrin}.

\subsection{CMB/LSS cosmological limits}

According to Big Bang cosmology, 
the masses of nonrelativistic neutrinos are related to the neutrino 
fraction of closure density by
$\sum_j m_j = 40\,\Omega_{\nu}\,h_{65}^2$~eV,
where $h_{65}$ is the present Hubble parameter in units of 65~km/(s~Mpc).
As knowledge of large-scale structure (LSS) formation has evolved,
so have the theoretically preferred values for the hot dark matter (HDM)
component, $\Omega_\nu$.  In fact, the values have declined.
In the once popular HDM cosmology, one 
had $\Omega_\nu \sim 1$ and $m_\nu \sim 10$~eV 
for each of the mass-degenerate neutrinos.
In the cold-hot CHDM cosmology, the cold matter was dominant 
and one had $\Omega_\nu\sim 0.3$ and $m_\nu \sim 4$~eV
for each neutrino mass.
In the currently favored $\Lambda$CDM cosmology with a non-vanishing 
cosmological constant $\Lambda$,
there is scant room left for the neutrino component.
The power spectrum of early-Universe density perturbations
is processed by gravitational instabilities.
However, 
the free-streaming relativistic 
neutrinos suppress the growth of fluctuations
on scales below the horizon 
(approximately the Hubble size $c/H(z)$) 
until they become nonrelativistic at redshifts 
$z\sim m_j/3T_0 \sim 1000\,(m_j/{\rm eV})$.

A recent limit \cite{elg} 
derived from the power spectrum obtained in the
2dF (Two Degree Field system) Galaxy Redshift Survey 
constrains the fractional contribution of massive neutrinos to the total
mass density to be less than 0.13
(for a total
matter mass density $0.1<\Omega_m<0.5$ and a scalar spectral index $n=1$). 
This  translates into a bound on the
sum of neutrino mass eigenvalues,  $\sum_j m_j<1.8$~eV.
Previous cosmological bounds from the data of galaxy cluster abundances,
the Lyman $\alpha$ forest and data compilations including the cosmic microwave 
background (CMB), peculiar 
velocities and large scale structure give upper bounds on the sum 
of neutrino masses in the range 3-6~eV (see also \cite{hannestad}).
It was estimated in \cite{hu} that the
precision determination of the  power spectrum shape by the Sloan Digital 
Sky Survey, combined with the CMB data of the MAP satellite 
experiment can reach a sensitivity of $\sum m_{\nu} \lsim 0.65$~eV.
For discussions of possible limits from future time of flight measurements
of supernova or gamma ray burst neutrinos see \cite{BBM00}.

Some caution is warranted in the cosmological approach to neutrino mass
in that the many cosmological parameters may conspire in 
various combinations to yield nearly identical CMB and LSS data.
An assortment of very detailed data may be needed to resolve 
the possible ``cosmic ambiguities''.

\section{Neutrinoless double beta decay}

\begin{figure}
\vspace{0.5cm}
\centerline{\resizebox{10cm}{5cm}{\includegraphics{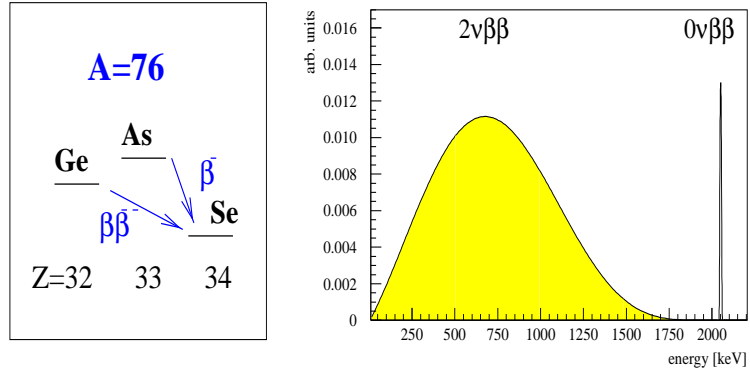}}}
 \caption{Left panel: Nuclear spectrum allowing for observable double beta 
decay. Right panel:
Energy spectrum for the SM allowed
$2\nu\beta\beta$ decay and $0\nu\beta\beta$ decay: the sharp peak at the 
Q-value provides a clear signature for non-vanishing neutrino masses}
 \label{znbb}
\end{figure}

The 
$0\nu\beta\beta$ \cite{0vbb} process
corresponds to two single beta decays 
occurring simultaneously in 
one
nucleus and 
converts a nucleus (Z,A) into a nucleus (Z+2,A), see Fig.~\ref{znbb}.
The SM allowed process emitting two antineutrinos,
\be{}
^{A}_{Z}X \rightarrow ^A_{Z+2}X + 2 e^- + 2 {\overline \nu_e},
\ee
is one of the rarest processes in nature with half lives in the region of
$10^{21-24}$ years. More interesting is the search for 
the neutrinoless mode,
\be{}        
^{A}_{Z}X \rightarrow ^A_{Z+2}X + 2 e^- 
\ee
which
violates lepton number by two units and thus implies physics beyond the 
SM. 

\begin{figure}
\vspace{0.5cm}
\centerline{\resizebox{5cm}{5cm}{\includegraphics{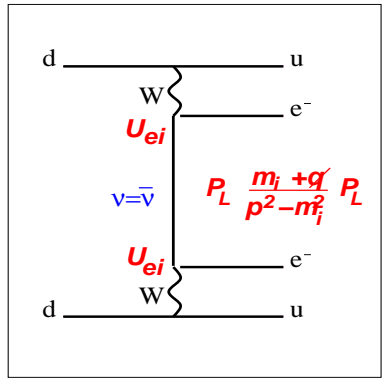}}}
 \caption{Diagram for neutrinoless double beta decay.}
 \label{diag}
\end{figure}

The $0\nu\beta\beta$ rate is a sensitive tool for the
measurement of the absolute mass-scale for Majorana neutrinos \cite{kps}.
The observable measured in the amplitude of $0\nu\beta\beta$ 
is the $ee$ element of the neutrino mass-matrix in the flavor basis
(see Fig.~\ref{diag}).
Expressed in terms of the mass eigenvalues and 
neutrino mixing-matrix elements, it is 
\be{}
m_{ee}= |\sum_i U_{ei}^2 m_i|\,.
\label{dbeqn}
\ee
A reach to a value as low as $m_{ee}\sim 0.01$~eV seems possible 
with proposed double beta decay projects such as 
GENIUSI, MAJORANA, EXO, XMASS or MOON. 
This provides a substantial improvement over the current bound of
$m_{ee}< 0.6$~eV. A recent report \cite{evi} 
of the Heidelberg-Moscow experiment
claims a best fit value of $m_{ee}=0.36$~eV,
but it is subject to controversions regarding the background determination.
In the far future,
another order of magnitude in reach 
is available to the 
10 ton version of GENIUS, provided it will
be funded and commissioned.

For masses in the interesting range $\gsim 0.01$~eV, 
the two light mass eigenstates are nearly degenerate and hence the 
approximation $m_1 =m_2$ (partial or total degeneracy) is justified.
Due to the restrictive CHOOZ bound, $|U_{e3}|^2 < 0.025$,
the contribution of the third mass eigenstate 
is nearly decoupled from $m_{ee}$ and hence
$U^2_{e3}\,m_3$ may be neglected in the $0\nu\beta\beta$ formula.
We label by $\phi_{12}$ the relative phase between
$U^2_{e1}\,m_1$ and $U^2_{e2}\,m_2$.
Then, employing the above approximations,
we arrive at a very simplified expression for $m_{ee}$:
\be{}
m^2_{ee}=\left[1-\sin^2 (2\theta_{\odot})\,
       \sin^2 \left(\frac{\phi_{12}}{2}\right)\right]\,m^2_1\,.
\label{dbeqn2}
%
\ee
The two CP-conserving values of $\phi_{12}$ are 0 and $\pi$.
These same two values give maximal constructive and destructive
interference of the two dominant terms in eq.\ (\ref{dbeqn}),
which leads to upper and lower bounds for the observable
$m_{ee}$ in terms of a fixed value of $m_1$:
\be{}
\cos (2\theta_{\odot})\;m_1 \leq m_{ee} \leq m_1 \,,
\quad {\rm for\;\;fixed}\;\;m_1\,.
\label{dbbnds}
\ee
The upper bound becomes an equality, $m_{ee}=m_1$, if $\phi_{12}=0$.
The lower bound depends on Nature's value of the mixing angle
in the LMA solution.
A consequence of eq.\ (\ref{dbbnds}) is that for a given
measurement of $m_{ee}$, the corresponding inference of $m_1$ is 
uncertain over the range 
$[m_{ee},\,m_{ee}\,\cos (2\theta_{\odot})]$
due to the unknown phase difference $\phi_{12}$, with 
$\cos (2\theta_{\odot}) > 0.1$ \cite{solanal}. 
This uncertainty disfavors $0\nu\beta\beta$
in comparison to direct measurements if a specific value of $m_1$
has to be 
determined, while $0\nu\beta\beta$ is more sensitive as long as 
bounds on $m_1$ are aimed at. 
Knowing the value of $\theta_{\odot}$ better will improve
the estimate of the inherent uncertainty in $m_1$.
The forthcoming KamLAND experiment should reduce the error in the 
mixing angle. 

\begin{figure}
\vspace{0.5cm}
\centerline{\resizebox{9cm}{7cm}{\includegraphics{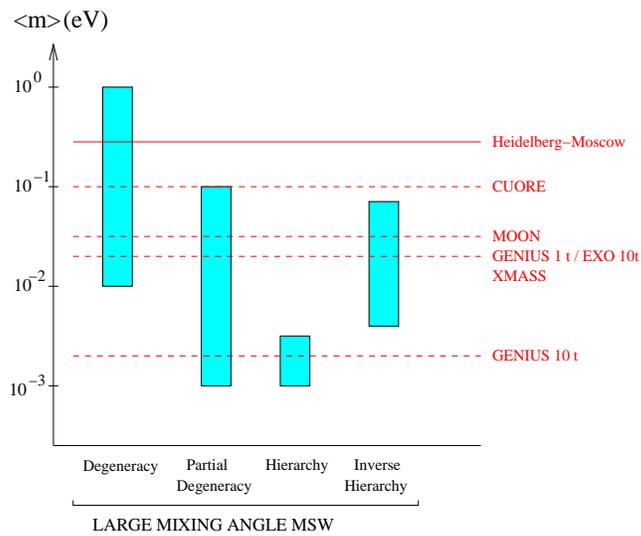}}}
 \caption{Different neutrino mass spectra versus sensitivities of future
double beta decay projects. A futuristic 10 ton Genius experiment may
test all neutrino spectra.
}
 \label{fig:cstates}
\end{figure}

A far future 10 ton version of GENIUS would be sensitive even to hierarchical
neutrino spectra, $m_1 \simeq 0 \ll m_2 \ll m_3$. A summary of
the
aimed sensitivities of future $0\nu\beta\beta$ projects to different neutrino
spectra is given in Fig.~\ref{fig:cstates}
including the inverse hierarchy, in which the $\nu_e$ admixture is
mainly in the heaviest state.

\section{Extreme energy cosmic rays in the Z-burst model}

It was expected that the extreme energy cosmic ray (EECR)
primaries would be protons from outside of our Galaxy, produced in
Nature's most extreme environments such as the tori or radio hot spots 
of active galactic nuclei (AGN).
Indeed, cosmic ray data show a spectral flattening just below
$10^{19}$~eV
which can be interpreted as a new extragalactic component overtaking
the lower energy galactic component;
the energy of the break correlates well with the onset of a Larmor radius 
for protons too large to be contained by the Galactic magnetic field.
It was further expected that the extragalactic spectrum 
would reveal an end at the 
Greisen-Kuzmin-Zatsepin (GZK) cutoff energy of
$\egzk \sim 5\times 10^{19}$~eV. 
The origin of the GZK cutoff is the degradation of nucleon energy by the 
resonant scattering process $N+\gamma_{2.7K}\rightarrow \Delta^*
\rightarrow N+ \pi$ when the nucleon is above the resonant threshold $\egzk$.
The concomitant energy-loss factor is
$\sim (0.8)^{D/6 {\rm Mpc}}$ for a nucleon traversing a distance $D$. 
Since no AGN-like sources are known to exist within 100
Mpc of the earth, the energy requirement for a proton arriving at the
earth with a
super-GZK energy is unrealistically high. 
Nevertheless, to date more than twenty events with energies 
at and above $10^{20}$~eV have been observed 
(for recent reviews see \cite{crrev}). \\

\begin{figure}
\vspace{0.5cm}
\centerline{\resizebox{6cm}{7cm}{\includegraphics{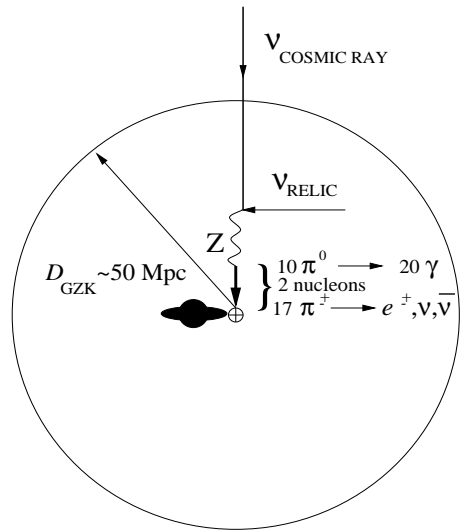}}}
 \caption{Schematic diagram showing the production of a Z-burst 
resulting from the 
resonant annihilation of a cosmic ray neutrino on a relic (anti)neutrino.
If the Z-burst occurs within the GZK zone ($\sim$ 50 to 100 Mpc) and is 
directed towards the earth, then photons and nucleons with energy above 
the GZK cutoff may arrive at earth and initiate super-GZK air-showers.}
 \label{fig:nu2}
\end{figure}

Several solutions have been proposed
for the origin of these EECRs,
ranging from unseen Zevatron accelerators (1~ZeV~$=10^{21}$~eV) 
and decaying supermassive particles and topological 
defects in the Galactic vicinity, 
to exotic primaries, exotic new interactions, and even exotic breakdown
of conventional physical laws.
A rather conservative and economical scenario involves cosmic ray 
neutrinos which scatter resonantly at the cosmic neutrino background (CNB) 
predicted by Standard Cosmology and produce Z-bosons \cite{Zburst}. 
These Z-bosons in turn decay to produce a highly boosted ``Z-burst'',
containing on average twenty photons and two nucleons above $\egzk$
(see Fig.~\ref{fig:nu2}).
The photons and nucleons from Z-bursts produced within a distance of
50 to 100 Mpc
of the earth can reach the earth with enough energy to initiate the 
air-showers observed at $\sim 10^{20}$~eV.

The energy of the neutrino annihilating at the peak of the Z-pole is
\be{}
E_{\nu_j}^R=\frac{M_Z^2}{2 m_j}=4\,(\frac{{\rm eV}}{m_j})\,{\rm ZeV}.
\ee
Even allowing for energy fluctuations about mean values, 
it is clear that in the Z-burst model the relevant
neutrino mass cannot exceed $\sim 1$~eV.
On the other hand, the neutrino mass cannot be too light. Otherwise
the predicted primary energies will exceed the observed
event energies and the primary neutrino flux will be pushed
to unattractively higher energies.
In this way,
one obtains a rough lower limit on the
neutrino mass of $\sim 0.1$~eV for the Z-burst model 
(with allowance made for an order of magnitude energy-loss 
for those secondaries traversing 50 to 100 Mpc). 
A detailed
comparison of the predicted proton spectrum with the observed EECR spectrum
in \cite{ringwald} yields a value of 
$m_{\nu}=0.26^{+0.20}_{-0.14}$~eV for extragalactic 
halo origin of the power-like part of the spectrum.

A necessary condition for the viability of this 
model is a sufficient flux of neutrinos at $\gsim 10^{21}$ eV.
Since this condition seems challenging, any increase of the Z-burst rate 
is helpful, 
that ameliorates slightly the formidable flux requirement.
If the neutrinos are mass degenerate, then a further consequence is that
the Z-burst rate at $E_R$ is three times what it would be 
without degeneracy. 
If the neutrino is a Majorana particle,
a factor of two more is gained in the Z-burst rate relative 
to the Dirac neutrino case since the two active helicity states 
of the relativistic CNB
depolarize upon cooling to populate all spin states.

Moreover the viability of the Z-burst model is enhanced if the CNB neutrinos 
cluster in our matter-rich vicinity of the universe.
For smaller scales, the Pauli blocking of identical
neutrinos sets a limit on density enhancement.
As a crude
estimate of Pauli blocking, one may use the zero temperature Fermi gas as a
model of the gravitationally bound neutrinos. Requiring that the Fermi
momentum of the neutrinos does not exceed the mass times the virial velocity 
$\sigma\sim\sqrt{MG/L}$ within the cluster of mass $M$ and size $L$, 
one gets the Tremaine-Gunn bound
\be{}
\frac{n_{\nu_j}}{54\,{\rm cm}^{-3}}\lsim 
10^3 \left(\frac{m_j}{{\rm eV}}\right)^3 
\left(\frac{\sigma}{200~{\rm km/s}}\right)^3\,.
\ee
With a  virial velocity within our Galactic halo 
of a couple hundred km/s,
it appears that Pauli blocking allows significant clustering on the
scale of our Galactic halo only if $m_j \gsim 0.5$~eV.
Free-streaming (not considered here) also works against HDM clustering.

Thus, if the Z-burst model turns out to be the correct 
explanation of EECRs, then
it is probable that neutrinos possess one or more masses in the range 
$m_{\nu}\sim (0.1-1)$~eV. 
Mass-degenerate neutrino models are then likely. 
Some consequences are:

\begin{itemize}

\item
A value of $m_{ee}>0.01$~eV, and thus
a signal of $0\nu\beta\beta$ in next generation experiments,
assuming the neutrinos are Majorana particles.

\item
Neutrino mass sufficiently large to affect the 
CMB/LSS power spectrum. 

\end{itemize}

\section{Conclusions}

Solar and atmospheric neutrino oscillations have established solid evidence 
for non-vanishing neutrino masses. 
A unique picture of the mixing matrix $U$ 
and mass squared differences in a three neutrino framework
is evolving with increasing accuracy.
Massive neutrinos also can provide information about the theoretical structures
underlying the standard model 
and may well be a key to the physics of supersymmetry, 
grand unification or extra dimensions.
A crucial if not the most important parameter in this context
is the absolute neutrino mass scale.
Information about absolute neutrino masses can be obtained from direct
determinations via tritium beta decay or cosmology. More sensitive in giving 
limits but less valuable for determining the mass scale is neutrinoless 
double beta decay. The most ambitious proposals for future double beta decay
experiments, such as the 10 ton version of GENIUS, may in fact test 
all possible neutrino mass spectra.
The puzzle of EECRs above the GZK 
cutoff can be solved conservatively
with the Z-burst model, connecting the ZeV scale of EECRs 
to the sub-eV scale
of neutrino masses. If the Z-burst model turns out to be correct, 
neutrino masses in the region of 0.1-1~eV are predicted and  
degenerate scenarios are favored.
In this case positive signals for future tritium beta decay experiments, 
CMB/LSS studies, and next-generation $0\nu\beta\beta$ experiments
can be expected.

\section*{Acknowledgements}
I would like to thank 
E. M\"uller-Hartmann and R. Schmitz for the kind invitation to
give a colloquium talk at the University of K\"oln this review is based on.
This work was supported by the Bundesministerium 
f\"ur Bildung und Forschung (BMBF, Bonn, Germany) under the 
contract number 05HT1WWA2.

\end{document}